\documentclass[11pt]{article}
\usepackage{moriond,epsfig}

\def\Journal#1#2#3#4{{#1} {\bf #2}, #3 (#4)}

\def\PR{\em Phys. Rev.}
\def\pr{\em Phys. Rep.}

\def\apj{ApJ}
\def\nat{\em Nature}
\def\aa{A\& A}
\def\mnras{MNRAS}
\def\proclon{Proc. Roy. Soc. London}

\def\be{\begin{equation}}
\def\ee{\end{equation}}
\def\bea{\begin{eqnarray}}
\def\eea{\end{eqnarray}}

\begin{document}
\vspace*{4cm}
\title{COSMIC SHEAR: THE DARK SIDE OF THE UNIVERSE}

\author{L. VAN WAERBEKE$^{1,2}$, Y. MELLIER$^{1,3}$, I. TERENO$^{1,4}$ }

\address{$^1$Institut d'Astrophysique de Paris\\
98bis Bd Arago, 75014, Paris, France\\
$^2$ Canadian Institute for Theoretical Astrophysics\\
60 St George Str., M5S 3H8, Toronto, Canada\\
$^3$ Observatoire de Paris, LERMA\\
61, avenue de l'observatoire, 75014, Paris, France\\
$^4$ Department of Physics, University of Lisbon\\
Campo Grande, Edificio C8, 1749-016 Lisboa, Portugal
}

\maketitle\abstracts{
We discuss the present status and future prospects for cosmic shear
observations and their cosmological constraints. We review the
evidences supporting the cosmological origin of the measured signal,
and discuss the possible problems coming from intrinsic alignment
and the actual limitations of theoretical predictions.
}

\section{Introduction and History}

The cosmic shear is a gravitational lensing effect which occurs
everywhere in the universe, and allows astronomers to map the
projected mass distribution on the sky from the solely observation
of the distorted distant galaxies. The idea of mapping the
matter using the gravitational deflection of ray-lights was born
in 1937 when F. Zwicky \cite{zwa,zwb} envisioned the possibility
to use the distorted shape of distant
galaxies to probe the matter content in nearby clusters of galaxies.
 His  idea was only discussed seriously
   30 years after, when first detailed
analytical work were produced \cite{ks,gunn},
motivated by the progress made in geometric optic in
curved spacetimes \cite{s61}.
But it is only in the early 90's that a robust link
was established between an appealing theoretical idea and the
observational possibilities \cite{me91,b91,k92}. In the meantime, 
  in 1983, a first attempt to measure cosmic shear failed \cite{v83} 
 mainly because of the poor image quality of data available at that time.
  The interest for cosmic shear raised again 
   after the discovery of giant arcs in 1987, and the burst of 
   theoretical papers started in the mid 90's
\cite{v96,b97,j97}, followed by many others \cite{bs01}.
 In 2000  the first detections
were reported almost simultaneously, and independently by four
teams \cite{b00,k00,vw00,w00}. Since then, several other measurements were done and
significant improvements in the data analysis lead to refined measures
and to the first robust cosmological constraints \cite{maoli01,vw01,rhodes01,pen02,vw02,hoekstra01,ham02,hoekstra02,bacon02,ref02}.

\section{Theory}

Gravitational lensing plays a special role in cosmology because it is the only
way to see the dark matter distribution from the galactic scale up to several
degrees. It is therefore the only observational tool which can measure directly
the mass power spectrum in the nearby universe with a direct link to
the constituents of the universe simultaneously in the linear and
non-linear dynamical regimes.

\begin{figure}
\centerline{
\psfig{figure=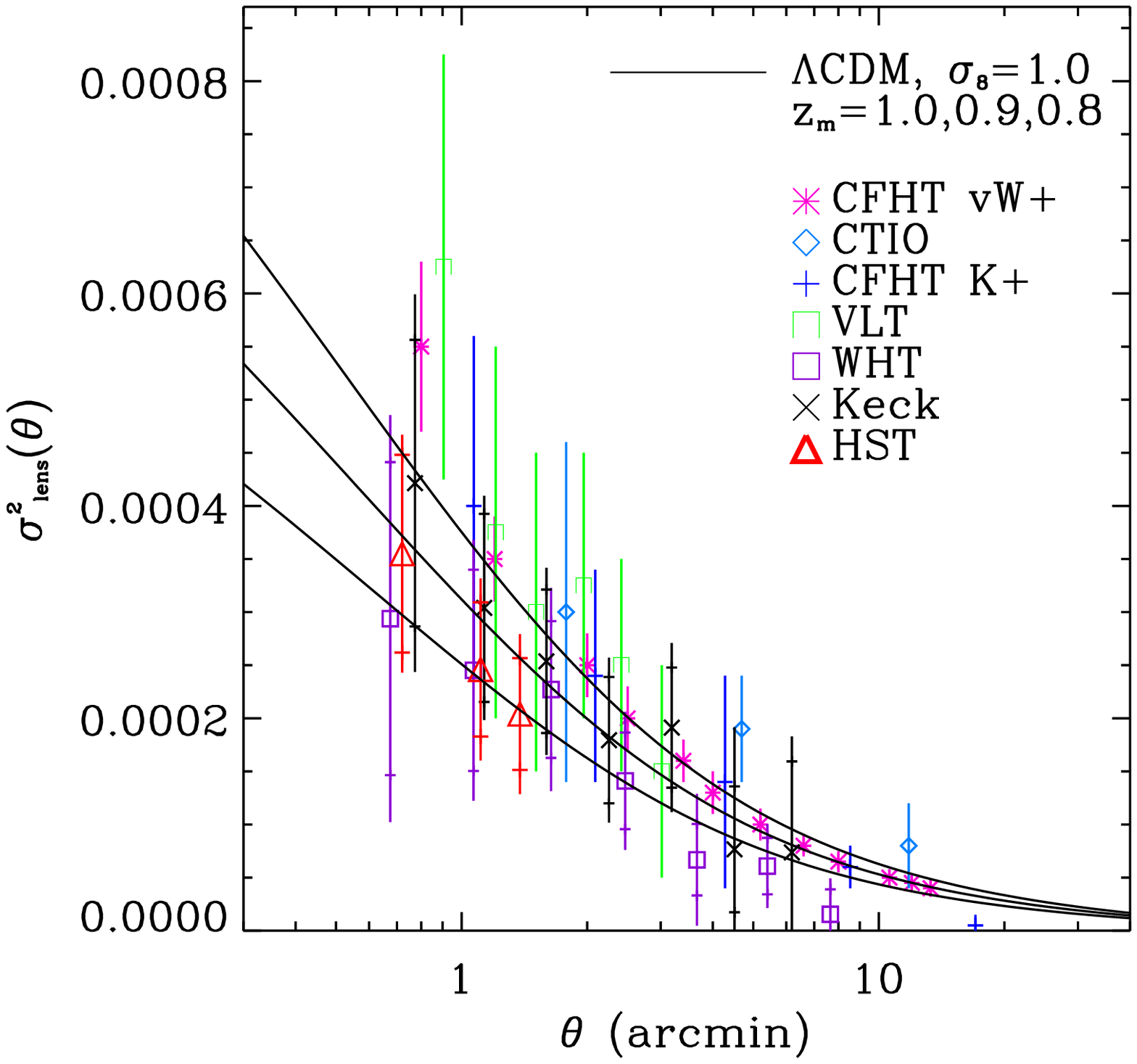,height=7cm}
\psfig{figure=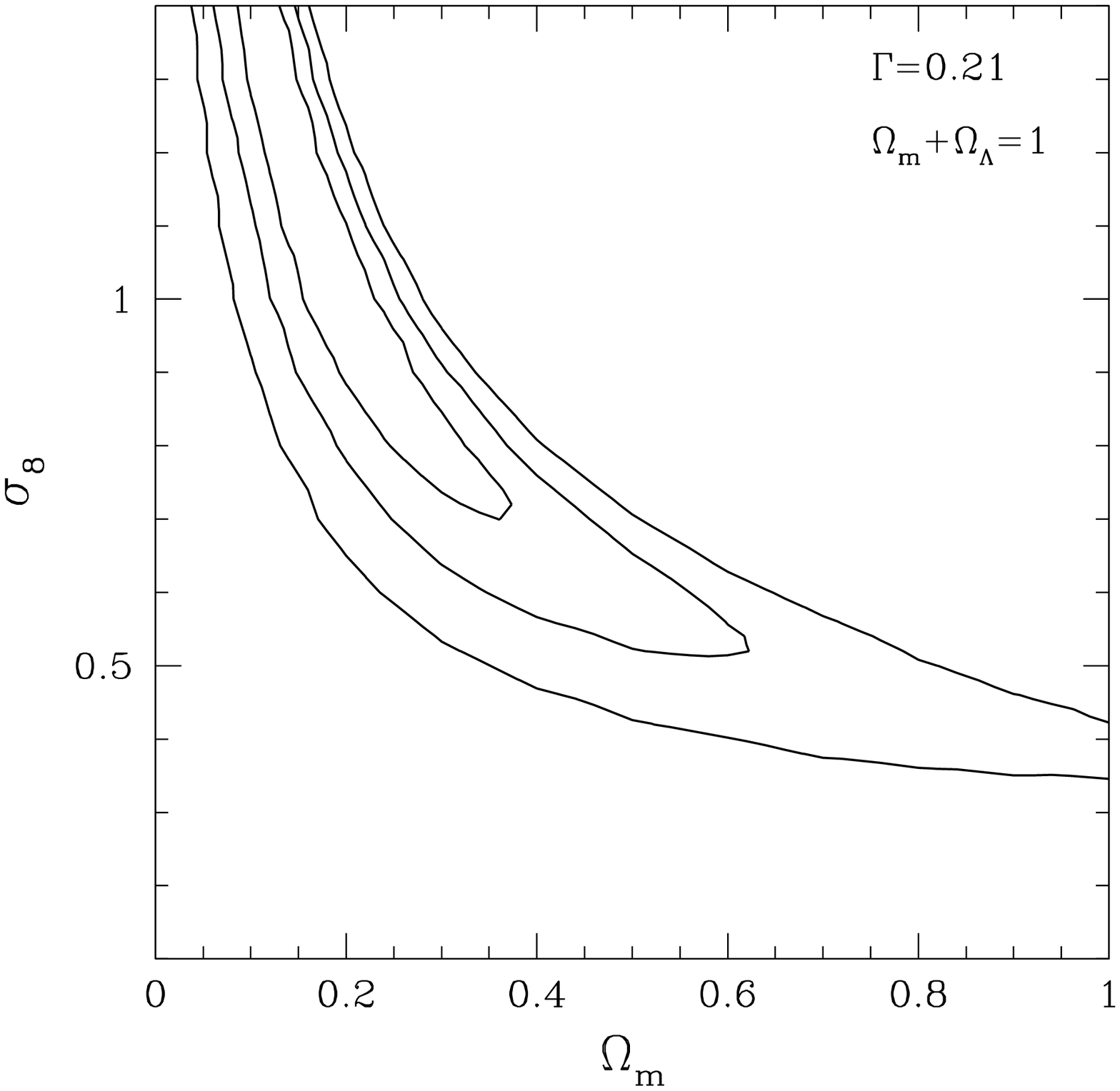,height=7cm}
}
\caption{Left: compilation of recent results of top-hat shear variance
measurements from several groups$^{31}$. Right: $\Omega_m$, $\sigma_8$
constraints for the Red Sequence Cluster Survey (RCS) from the shear
top-hat variance measurements$^{17}$.
\label{fig:tophatstat}}
\end{figure}

The power spectrum of the projected mass, called the convergence power spectrum
$P_\kappa(k)$, is the quantity which relates any cosmic shear two points statistics
to the cosmological parameters and the 3-dimensional mass power
spectrum $P_{3D}(k)$:

\begin{equation}
P_\kappa(k)={9\over 4}\Omega_0^2\int_0^{w_H} {{\rm d}w \over a^2(w)}
P_{3D}\left({k\over f_K(w)}; w\right)
\left[ \int_w^{w_H}{\rm d} w' n(w') {f_K(w'-w)\over f_K(w')}\right]^2,
\label{pofkappa}
\end{equation}
where $f_K(w)$ is the comoving angular diameter distance out to a
distance $w$ ($w_H$ is the horizon distance), and $n(w(z))$ is the
redshift distribution of the sources. The mass power spectrum
$P_{3D}(k)$ is evaluated in the non-linear regime
\cite{pd96}, and $k$ is the 2-dimensional wave vector
perpendicular to the line-of-sight. The three most common
   observables are respectively the
shear top-hat variance \cite{me91,b91,k92}, the aperture mass
variance \cite{k94,sch98} and the shear correlation
function \cite{me91,b91,k92}:

\begin{equation}
\langle\gamma^2\rangle={2\over \pi\theta_c^2} \int_0^\infty~{{\rm d}k\over k} P_\kappa(k)
[J_1(k\theta_c)]^2,
\label{theovariance}
\end{equation}
\begin{equation}
\langle M_{\rm ap}^2\rangle={288\over \pi\theta_c^4} \int_0^\infty~{{\rm d}k\over k^3}
P_\kappa(k) [J_4(k\theta_c)]^2,
\label{theomap}
\end{equation}
\begin{equation}
\langle\gamma(r)\gamma(r+\theta)\rangle_r={1\over 2\pi} \int_0^\infty~{\rm d} k~
k P_\kappa(k) J_0(k\theta),
\label{theogg}
\end{equation}
where $J_n$ is the Bessel function of the first kind. They are all different
measurements of the same physical quantity, the convergence power spectrum
$P_\kappa(k)$. Their internal
consistency provides a valuable check of the cosmological origin of the observed
signal.

\section{Measurements and cosmological constraints}

\begin{figure}
\centerline{
\psfig{figure=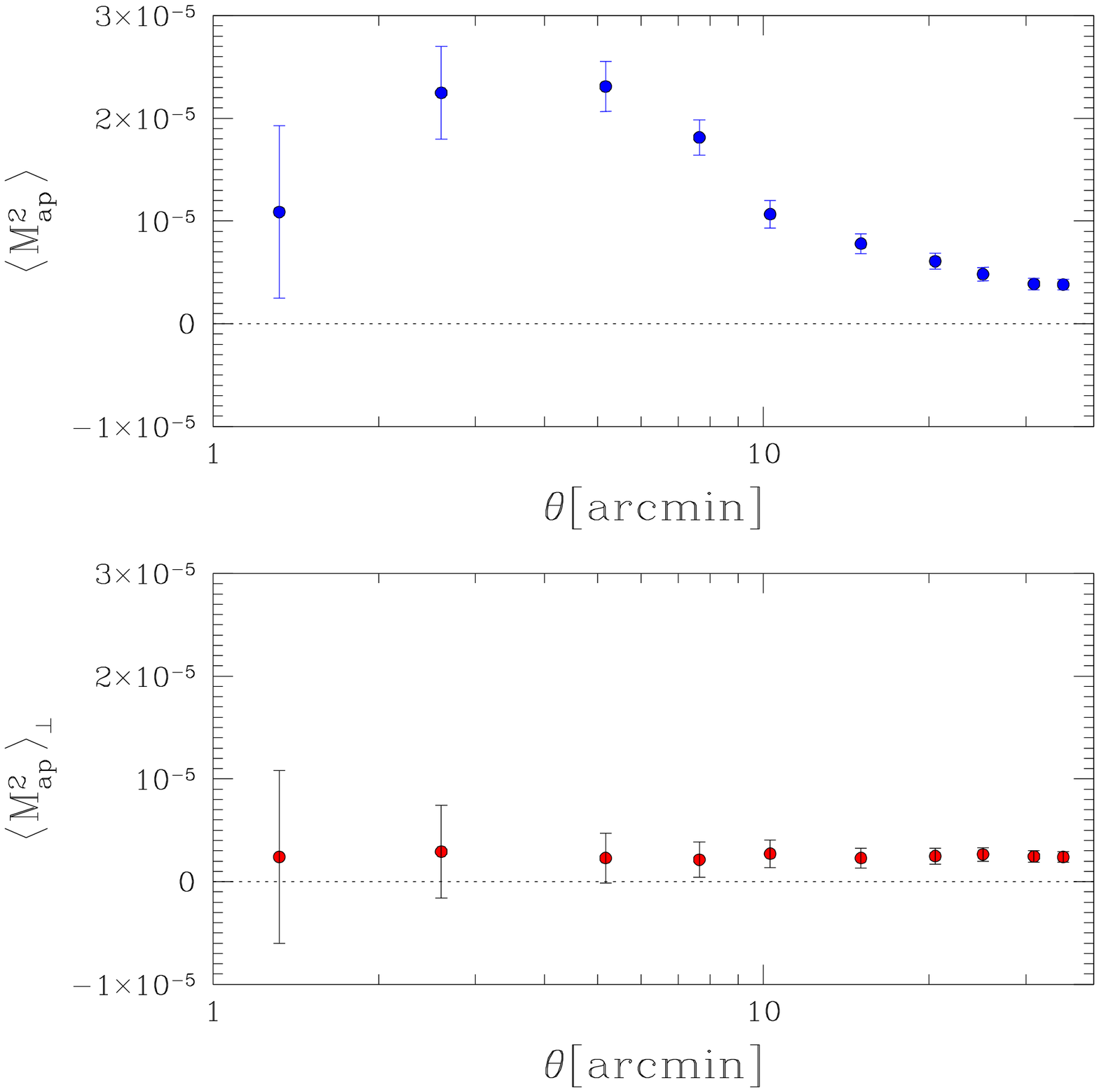,height=7cm}
\psfig{figure=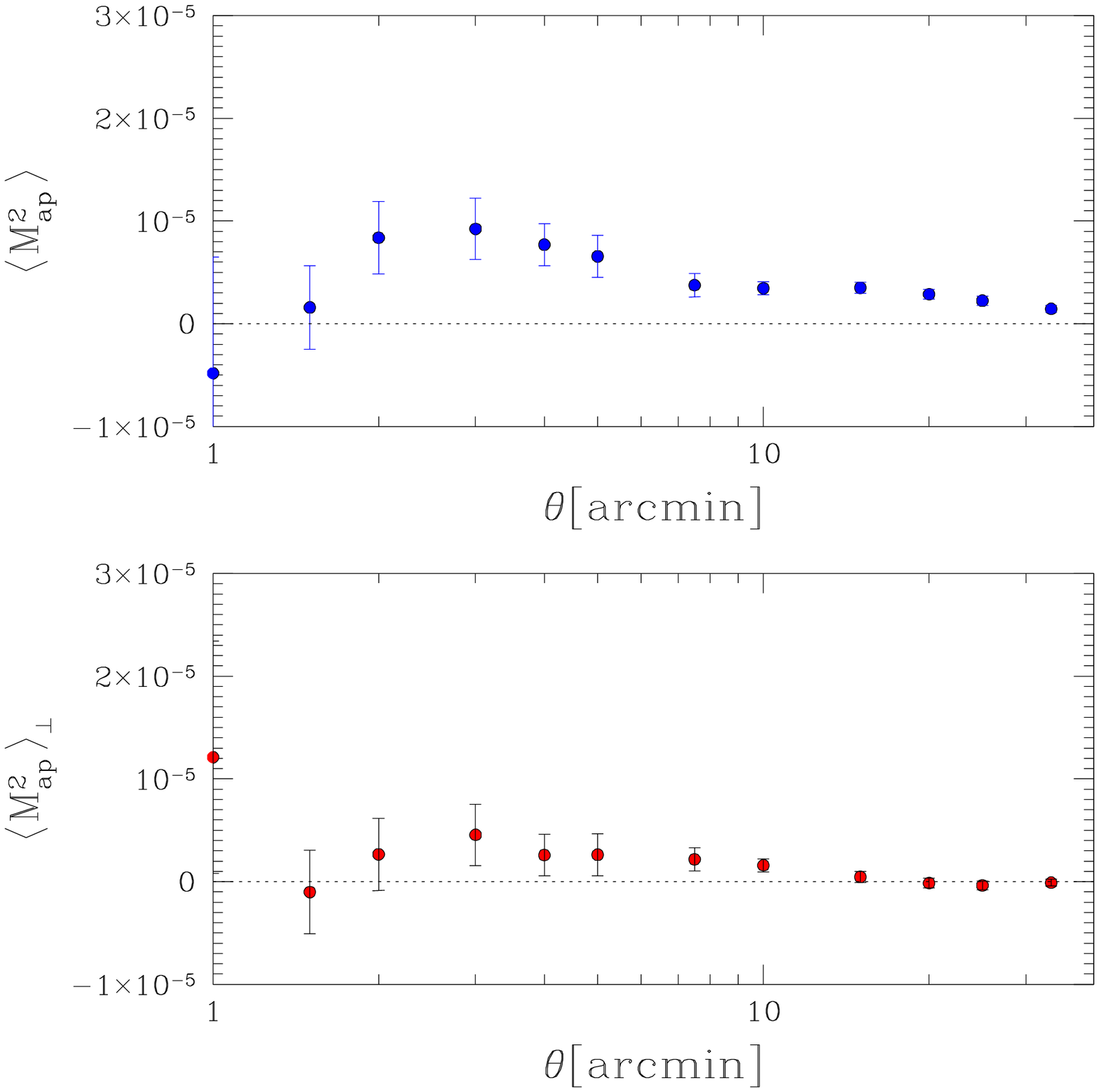,height=7cm}}
\caption{Left: $E$ (top) and $B$ (bottom) modes measured in the VIRMOS
survey$^{39}$. Right: $E$ (top) and $B$ (bottom) modes measured in the
RCS survey$^{16}$. The $B$ mode is low and the $E$ mode compatible
with the predictions for the aperture mass statistics$^{34}$.
\label{fig:mapstat}}
\end{figure}

As quoted earlier \cite{j97}, the cosmic shear signal depends primarily
on four parameters: the cosmological mean density $\Omega_m$, the mass power
spectrum normalisation $\sigma_8$, the shape of the power spectrum $\Gamma$,
and the redshift of the sources $z_s$. Therefore any measurement of the
statistics shown in Section 2 could provide constraints on these parameters.
Left plot on Figure \ref{fig:tophatstat} shows a compilation
of the measurements of the
top-hat variance done by different groups, compared with a typical $\Lambda$CDM
model. The agreement of these measurements, done by different people, and
with different telescopes, support the idea of cosmological origin of the signal.
The right plot on Figure \ref{fig:tophatstat} shows the constraints
on $\Omega_m$ and $\sigma_8$
obtained from the top-hat variance of the shear measured in
the Red Sequence Cluster survey \cite{hoekstra01}. Strong assumption have been
applied to the hidden parameters $\Gamma$ and the source redshift,
which allows to break the degeneracy between
$\Omega_m$ and $\sigma_8$ (see also \cite{vw01}).  
However, the top-hat variance is a rather weak statistics, in the sense it does
not provide direct checks of possible contamination by systematic effects
(like residual optical distortion and telescope tracking, or intrinsic
alignment). For this reason,
various tests have been proposed and developed to provide robust checks
of the lensing effects, and to quantify the cosmological origin of the
signal. They are the following:

-The $E$, $B$ modes decomposition separates the lensing signal into
curl and curl-free modes \cite{crit02}. It is expected, and it can also
be quantified on the star field, that residual systematics equally
contributes to $E$ and $B$, while the lensing signal should be present ONLY
in the $E$ mode because gravity derives from a true scalar field \cite{steb96}.
The $E$ mode is identical to the aperture mass statistics Eq.(\ref{theomap})
\cite{k94,sch98}. The $E$ and $B$ modes have been measured in several surveys
\cite{vw01,vw02,pen02,hoekstra02}, and support the cosmological origin of the
signal, as well as showing the already small amount of residual systematics
achieved with today's technology. Figure \ref{fig:mapstat} shows such measurements
for the VIRMOS \footnote{http://www.astrsp-mrs.fr}
\footnote{http://terapix.iap.fr/DESCART} and RCS
\footnote{http://www.astro.utoronto.ca/~gladders/RCS/} surveys.

-The self-consistency of the three statistics shown in Section 2 was shown
for the VIRMOS \cite{vw01,pen02} and RCS surveys \cite{hoekstra01,hoekstra02}.

-The lensing signal is expected to decrease for low redshift sources, just
because the gravitational distortion becomes less efficient. This decrease
of the signal has been observed recently for the first time, when comparing
the VIRMOS survey aperture mass \cite{vw02} which has a source mean redshift
around $0.9$ to the RCS which has a source mean redshift around $0.6$. The
expected decrease in signal amplitude is about $2$, which is what is observed
(see Figure \ref{fig:mapstat}).

-Space images provide in principle systematics-free environment, and even if
the observed areas are still smaller than ground based observations, space
data provide ideal calibrations of the cosmic shear signal
\cite{rhodes01,ham02,ref02},
which are in excellent agreement with ground based measurements.

\begin{figure}
\centerline{\psfig{figure=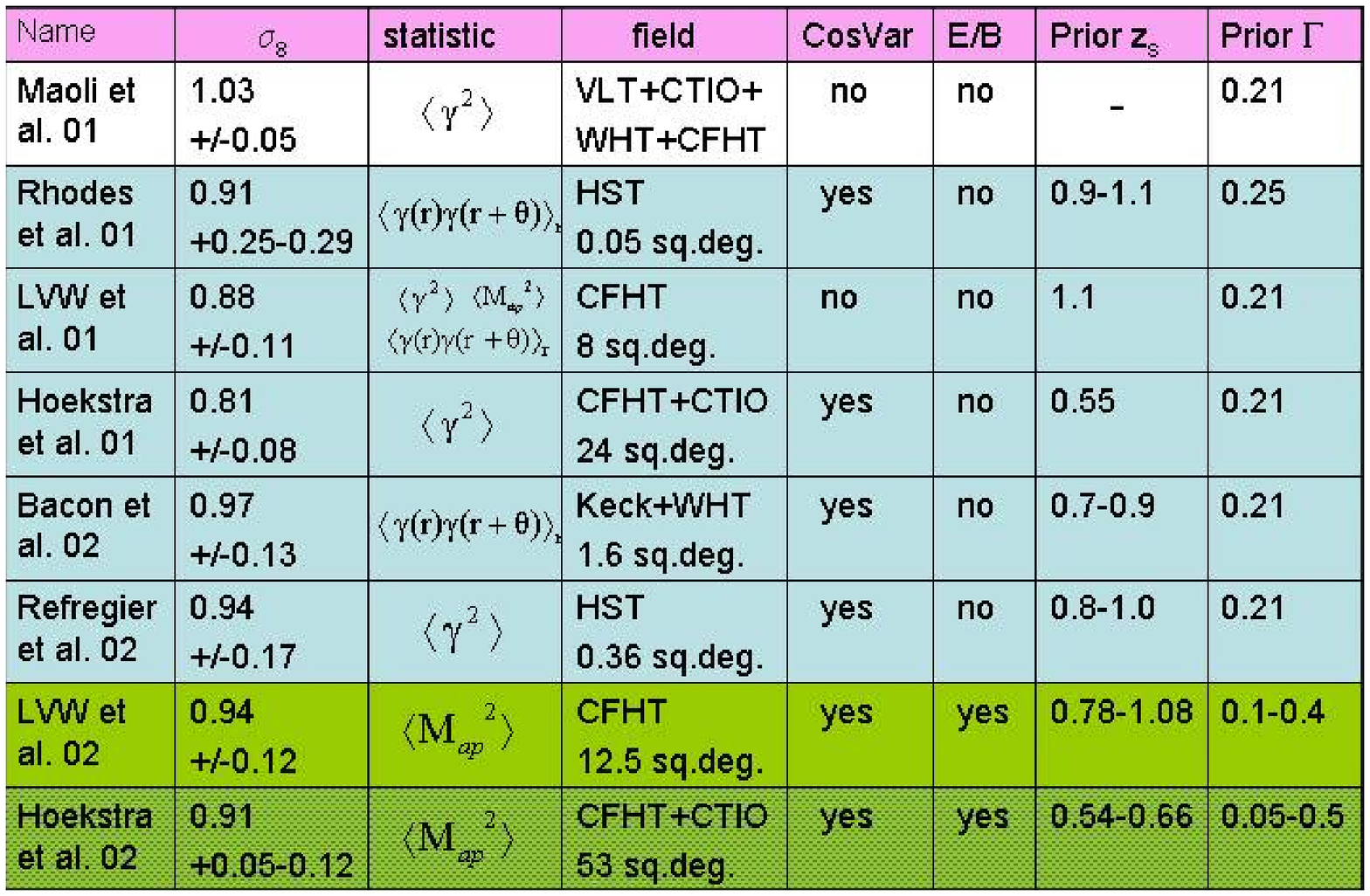,height=10cm}}
\caption{Constraints on $\sigma_8$ for a flat cosmology with $\Omega_m=0.3$.
The meaning of the different columns are: statistics; which statistic is used (see
Section 2), field; telescope used and area observed, CosVar; whether or not
cosmic variance has been included in the error bars, E/B; whether or not the
$E$, $B$ mode separation has been used, Prior $z_s$; flat prior on the mean
source redshift applied, Prior $\Gamma$; flat prior on $\Gamma$ applied.
\label{fig:constraints}}
\end{figure}

Figure \ref{fig:constraints} shows the constraints on $\sigma_8$ obtained for
a flat cosmology with $\Omega_m=0.3$ for various surveys, statistics and priors.
There is an overall good agreement, and some preference is given
for a normalisation slightly smaller than $1$ (a comparison with cluster
normalisation results is given elsewhere \cite{vw02}).

\section{Comments}

Gravitational lensing is not the only natural process which produces alignment of
galaxies over large distances. Intrinsic alignment might occur from tidal
fields, and produce galaxy shape correlations over cosmological distances, and
contaminate cosmological signal \cite{crometz,catelan,heav00,pen00,catelan01,hat01},
which should
in principle split, in a predictable way, into $E$ and $B$ modes. There is
unfortunately only partial agreement between these different
approaches. Yet it is hard to have a robust prediction for
the intrinsic alignment effect, although it is not believed to be higher than
a $10\%$ contribution for a lensing survey with a mean source
redshift at $z_s=1$. A recent work \cite{jing} suggested that intrinsic alignment
could dominate the cosmic shear even in deep surveys, which seems
incompatible with the observations:
this would indeed imply a very low $\sigma_8 \sim 0.1$ and we should also observe
an increase of the effect as we go from deep to shallow survey, which is not
the case (see Figure \ref{fig:mapstat}). In any case, intrinsic alignment
contamination can be removed completely by measuring the signal correlation
between distant redshift bins, instead of measuring the full projected signal.

Another issue is the cosmological parameter determination, which relies on
the accuracy of the non-linear
prediction of the cosmic shear signal. The non-linear predictions \cite{pd96}
are only accurate to $\sim10\%$ \cite{vw02}, which limitates the measurement
of cosmological parameters to the same accuracy. Therefore a net progress on
non-linear predictions will have to be made before cosmic shear measurements
can provide high-precision constraints on the parameters.

\section{The Future}

\begin{figure}
\centerline{\psfig{figure=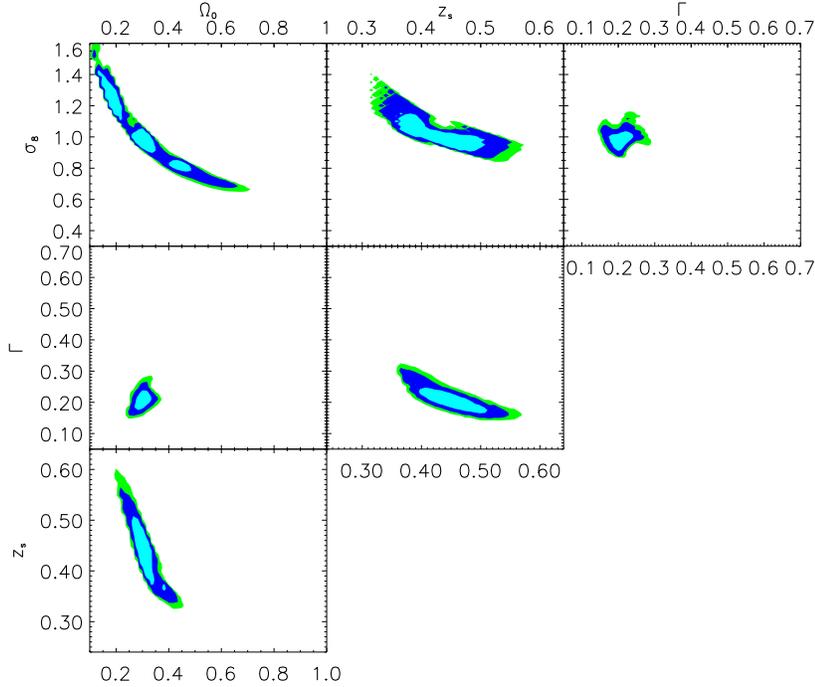,height=9cm}}
\caption{$1$, $2$ and $3\sigma$ contours for the parameters $\Omega_m$,
$\sigma_8$, $\Gamma$ and $z_s$, assuming a $130~{\rm sq. deg.}$ survey
and $m_{I_{AB}}=24.5$.
Constraints are obtained from the measurement
of the shear correlation function from $0.6'$ to $30'$. On each plot, hidden
parameters are marginalised as defined in the text. The true model is
$\Omega_m=0.3$, $\sigma_8=1.$, $\Gamma=0.21$ and $z_s=0.44$.
Note that, because of the chosen parametrisation, the mean redshift
of the survey is $2z_s$.
\label{fig:cfhtls_param}}
\end{figure}

Over the next $5$ years, the Canada France Hawaii Telescope Legacy Survey (CFHTLS)
will produce $130~{\rm sq. deg.}$ of high image quality data down to
$m_{I_{AB}}=24.5$. The observations in $5$ colors will allow a good determination
of photometric redshifts, and the large area will provide accurate measurements
of the projected mass power spectrum from $l=90$ to $l=10^5$. Procided that
residual systematics can be eliminated, accurate
cosmological parameter determination will
be possible, and the joint analysis with other experiments
(CMB, 2dF, SLOAN) will break parameter degeneracies
(Van Waerbeke et al. in preparation).
Figure \ref{fig:cfhtls_param} is an example of constraints on $\Omega_m$,
$\sigma_8$, $\Gamma$ and $z_s$, where the hidden parameters of each plot
have been marginalised using the priors $\Omega_m\in [0.25,0.35]$,
$\sigma_8\in[0.95,1.05]$, $\Gamma\in [0.1,0.3]$, $z_s\in[0.4,0.5]$. The true model
is $\Omega_m=0.3$, $\sigma_8=1.$, $\Gamma=0.21$ and $z_s=0.44$ (the
mean source redshift is $2z_s$). Error bars contain gaussian cosmic variance
\cite{sch02} and statistical noise corresponding to
$m_{I_{AB}}=24.5$ \cite{vw02}. It is clear that accurate constraints can be
obtained, but it is also very important to estimate the
redshift distribution, as Figure \ref{fig:cfhtls_param} shows
that the redshift axis is always strongly degenerate.

New roads have also been recently opened. The shear three point function
has probably been detected \cite{b02}, which potentially is a direct
measure of $\Omega_m$ \cite{b97}. The very first measurement
of the bias from cosmic shear has also been claimed \cite{hhg,hvw02}
which paved the way for galaxy formation studies from cosmic shear.


\end{document}